\documentclass{article}
\usepackage{spconf,amsmath,epsfig}
\usepackage{graphicx}
\usepackage{subfigure}
\usepackage{multirow}
\usepackage{color}
\usepackage{changepage}
\usepackage{times}
\usepackage{float}
\usepackage{amssymb}
\usepackage{multirow}
\usepackage{bbding}
\usepackage{booktabs}
\usepackage{caption}
\usepackage{cite}
\usepackage{stfloats}
\usepackage{url}
\begin{document}\sloppy

\def\x{{\mathbf x}}
\def\L{{\cal L}}

\title{Portable health screening device of respiratory infections}
%

\name{Zheng Jiang$^{1}$, Menghan Hu$^{2}$, Guangtao Zhai$^{1}$\thanks{This work is sponsored by the National Natural Science Foundation of China (No. 61901172, No. 61831015, No.  U1908210), the Shanghai Sailing Program (No.19YF1414100), and the Science and Technology Commission of Shanghai Municipality (No. 19511120100).
}}
\address{$^1$Institute of Image Communication and Information Processing, Shanghai Jiao Tong University, China\\
	$^2$Shanghai Key Labora. of Multidim. Infor. Proce., East China Normal University, China}
\maketitle

\begin{abstract}

The COVID-19 epidemic was listed as a public health emergency of international concern by the WHO on January 30, 2020. To curb the secondary spread of the epidemic, many public places were equipped with thermal imagers to check the body temperature. However, the COVID-19 pneumonia has concealed symptoms: the first symptom may not be fever, and can be shortness of breath. During epidemic prevention, many people tend to wear masks. Therefore, in this demo paper, we proposed a portable non-contact healthy screening system for people wearing masks, which can simultaneously obtain body temperature and respiration state. 
This system consists of three modules viz. thermal image collection module, health indicator calculation module and health assessment module. In this system, the thermal video of human faces is first captured through a portable thermal imaging camera. Then, body temperature and respiration state are extracted from the video and are imported into the following health assessment module. Finally, the screening result can be obtained.
The results of preliminary experiments show that this syetem can give an accurate screening result within 15 seconds. This system can be applied to many application scenarios such as community and campus. The demo videos of the proposed system are available at: \url{https://doi.org/10.6084/m9.figshare.12028032}.
\end{abstract}
\section{Introduction}
\label{sec:intro}

Under the epidemic situation of COVID-19, the portable health screening device can effectively assist relevant departments in epidemic control. Some respiratory infectious diseases such as COVID-19 have hidden symptoms, and the first symptom in many cases is not fever, but shortness of breath. 
For these infectious diseases, leveraging these symptoms to find infected people can stop the virus from spreading again and allow infected people to get timely treatment. At present, the single health portrait such as body temperature constructed by the existing screening technology has great limitations such as low detection rate and high missing rate. 

The breathing state assessment is not mentioned as much as the body temperature in epidemic control. Respiration is one of the most significant characteristics of human vital signs. Among all the characteristics of respiration, respiratory rate and breathing patterns are particularly important. Studies illustrate that their changes are one of the earliest and major signs of many serious illnesses. 
So far, diverse studies have demonstrated that thermal imaging can be used to analyze the respiratory rate and breathing patterns\cite{hu2018influence}\cite{chen2019rgb}.
Hu et al. used thermal and visible imaging to achieving the noncontact and nonobtrusive measurements of breathing rate and pattern\cite{hu2017synergetic}.
\begin{figure}[h]
	\centering
	\includegraphics[width=0.4\textwidth]{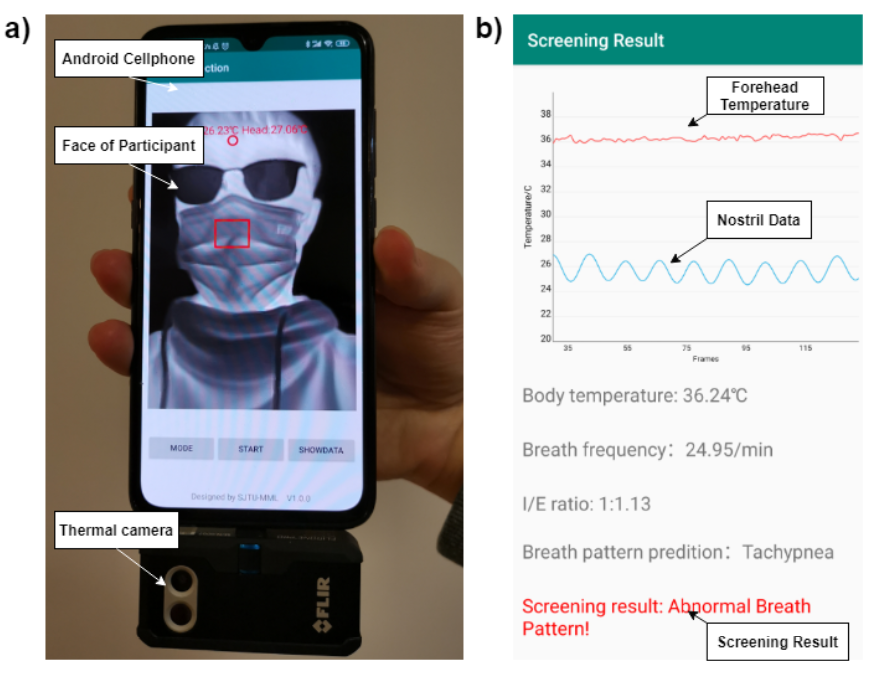}
	\caption{Overview of portable health screening system for respiratory infections: a) device appearance; b) analysis result of the application.}\label{F01}
\end{figure}

\begin{figure*}[ht]
	\centering
	\includegraphics[width=0.85\textwidth]{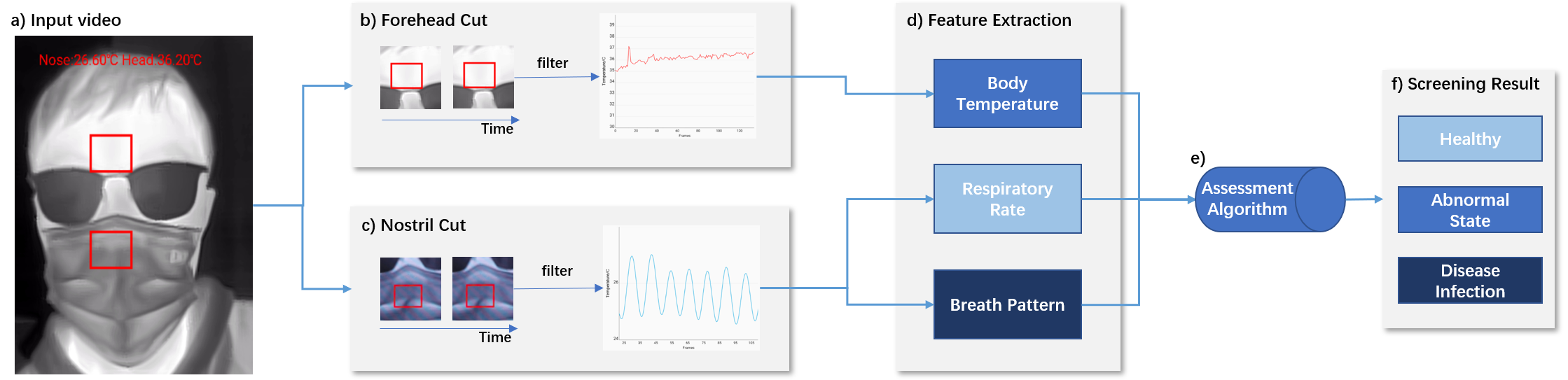}
	\caption{Pipeline for the health screening system: a) the input facial thermal video of participants; b) and c) capture the forehead and nostril cut of the video seperately and transfer images to different signals; d) extract different features according to the indicators; e) use a classfiy algorithm to perform the health assessment result; f) screening result.}\label{F02}
\end{figure*}
Therefore, we developed a remote respiratory infectious screening system based on portable thermal camera to help fight the current epidemic caused by SARS-CoV-2. The structure of our  screening system is shown in Fig. \ref{F01}.
It has the advantages of portability, low cost and easy operation. Through the 15-second detection window, this system can accurately determine body temperature and respiration state of people wearing masks, thus giving the screening result.


The main contributions of this demonstration are threefold. First, we combine the face recognition technology with portable thermal camera to accomplish the nostril recognition of facial thermal images. Subsequently, the temperature information and respiration state is successfully obtained by applying image analysis algorithms to facial thermal images. Finally, based on the existing equipment conditions, we have implemented a non-contact and efficient measurement system for health screening on respiratory infections.



\section{Demonstration Setup}
In our demonstration, we use a FLIR one thermal camera connected to an Android phone to collect thermal video, and use the application we developed to achieve real-time health screening of respiratory infections. The pipeline for our screening system is shown in Fig. \ref{F02}.
Three modules are inserted in our developed application, which are thermal image collection module, health indicator calculation module and health assessment module. 
The detailed descriptions of these three modules are presented below:

1) Thermal image collection module: this module is mainly composed of hardware devices, which consists of an Android smartphone and a FLIR one thermal camera. The system can be put into use simply by connecting the camera to the phone and installing the application we developed. In this demonstration, a thermal camera is aimed at the face of a participant wearing a mask to obtain a corresponding facial thermal video.


2) Health indicator calculation module: this module is mainly responsible for extracrting features of temperature and respiration data from the thermal videos. When human body performs continuous periodic breathing activities, continuous heat exchange occurs between the nostril and the air, and periodic temperature fluctuations occur in the respiratory tract. Therefore, respiration data can be obtained by analyzing the periodic changes in the temperature around the nostril. After using a series of algorithms to deal with the collected temperature data, the accurate health indicator can be obtained. In this system, we extract respiratory rate and one of the four common breathing patterns (Eupnea, Bradypnea, Tachypnea and Apnea) according to the obtained nostril signals. Body temperature is achieved through forehead signals.


3) Health assessment module: this module includes health screening and abnormal alert. Health screening is accomplished through a comprehensive analysis of body temperature, respiratory rate and breathing pattern. The algorithm is based on Hu's research\cite{hu2017synergetic}. Besides, an alert would be provided when abnormal health conditions are found.


\section{Visitors Experience}
In our experiments, participants wearing masks only need to stay hold in front of the thermal camera for 15 seconds to complete the health screening. The result shows that the system can efficiently extract the temperature and respiration data of participants, and give an accurate assessment of health conditions. During the whole non-contact screening process, slight movement of participants can be accepted.
This system may have great application prospects in many scenarios such as community, campus and hospital which can be used to perform preliminary check for respiratory infections and give abnormal alerts.



{
\bibliographystyle{IEEEbib}
\small
\bibliography{IEEEabrv,refs}

\begin{thebibliography}{1}

\bibitem{hu2018influence}
Menghan Hu, Guangtao Zhai, Duo Li, Hanqi Li, Mengxin Liu, Wencheng Tang, and
  Yuanchun Chen,
\newblock ``Influence of image resolution on the performance of remote
  breathing rate measurement using thermal imaging technique,''
\newblock {\em Infrared Physics \& Technology}, vol. 93, pp. 63--69, 2018.

\bibitem{chen2019rgb}
Lushuang Chen, Ning Liu, Menghan Hu, and Guangtao Zhai,
\newblock ``{RGB}-thermal imaging system collaborated with marker tracking for
  remote breathing rate measurement,''
\newblock in {\em 2019 IEEE Visual Communications and Image Processing (VCIP)}.
  IEEE, 2019, pp. 1--4.

\bibitem{hu2017synergetic}
Meng-Han Hu, Guang-Tao Zhai, Duo Li, Ye-Zhao Fan, Xiao-Hui Chen, and Xiao-Kang
  Yang,
\newblock ``Synergetic use of thermal and visible imaging techniques for
  contactless and unobtrusive breathing measurement,''
\newblock {\em Journal of biomedical optics}, vol. 22, no. 3, pp. 036006, 2017.

\end{thebibliography}
}

\end{document}